\documentclass{kapproc}
  \kluwerbib

\begin{document}

\articletitle{Distance
Determination of Variable Galactic Sources}

\author{}
\author{J. Hu$^1$, S. N. ZHANG$^{1,2,3,4}$ and T.P. Li$^{1,4}$}
\affil{$^1$Physics Department and Center for Astrophysics, Tsinghua
University, Beijing 100084, P.R. China}
\affil{$^2$Physics
Department, University of Alabama in Huntsville, Huntsville,
AL 35899, USA}
\affil{$^3$Space Science Laboratory,
NASA Marshall Space Flight Center, SD50, Huntsville, AL 35812, USA}
\affil{$^4$Institute of High Energy Physics, Chinese
Academy of Sciences, Beijing, China}

\begin{abstract}

We have developed a timing analysis method to determine the
distances of variable galactic X-ray sources based on the method
advanced by Tr\"{u}mper and Sch\"{o}nfelder in 1973. The
light-curve of the halo produced by the scattering of X-rays off
the interstellar dust is delayed and smeared by the dust grains.
This method utilizes the differences between the power density
spectra of the point source and the halo. We present the details
of this method and our first applications of this method to the
Chandra data of X-ray binary Cyg X-3.

\end{abstract}


\section{Introduction}

When the high energy photons emitted by the X-ray source arrive to
the earth, they have been scattered by the interstellar dust
grains. Therefore many galactic X-ray sources are observed to have
 faint and diffuse halos. This effect and its possible
application was first discussed by Overbeck (1965), and the X-ray
halo was first detected with Einstein Observatory (Rolf 1983,
Catura 1983). The investigation of the shape and strength of the
haloes was done by several authors (e.g. Mauche \& Gorenstein
1986, Predehl \& Schmitt 1995) based on the data of Einstein and
ROSAT.

As the photons in the halo travel through longer paths than the
photons in the point source, any intensity variation in the the
source induces a delayed signal in the halo. It has been suggested
by Tr\"{u}mper and Sch\"{o}nfelder (1973, hereafter TS1973) that
this effect can be uesd to determine the distance of variable
galactic X-ray sources. Although the method is very simple, the
first successful application was realized by Predehl (2000) for a
X-ray binary Cyg X-3 almost 30 years later. They obtained the
distance of $9^{+4}_{-2}$ kpc to the source. The difficulty of the
method is that it is hard to distinguish the halo from the image
induced by the point spread function of the instrument until the
Chandra X-ray Observatory provides us the high resolution image of
X-ray sources.

The shape and strength of the halo depend on the distance to the
X-ray source, the geometric distribution and the physical
properties of the interstellar dust grains (Hayakawa (1970),
Mathis \& Lee (1991), and Predehl \& Klose (1996)).

\section{Methodology}

The differential scattering cross section of X-ray photons in
interstellar dust is (Predehl, 2000)
\begin{equation}
\frac{{\rm d}\sigma}{{\rm
d}\Omega}=|S(\Phi)|^2\propto\exp(-0.46E^2a^2\Phi^2)
\end{equation}
where $a$ is the size of the dust grains (in $\mu$m), $E$ is the
energy of X-ray (in keV), and $\Phi$ is the scattering angle (in
arc min).

\begin{figure}[htb]
\begin{center}
\vskip 1in \includegraphics{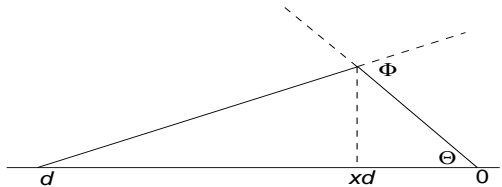} \caption {Scattering geometry. 0 indicates the
observer, $d$ indicates the source, $xd$ is where the scattering
occurs.}
\end{center}
\end{figure}

Following TS1973, we obtain the time delay of a scattered photon
compared with an unscattered photon is (See Fig. 1)
\begin{equation}\label{syj}
t=\frac{d}{2c}\frac{x\Theta^2}{1-x}=1.15d\frac{x\Theta^2}{1-x},
\end{equation}
where $d$ is given in kpc, $\Theta$ in arcsec.

Given the dust density distribution along the line of sight
$\rho(x)$, the time delay distribution function $H(t,\Theta)$ can
be calculated. As pointed by Predehl \& Schmitt (1995), we can
assume $\rho(x)$ is a constant. Thus
\begin{equation}
H(t,\Theta){\rm d}t{\rm d}\Theta=\rho(x)S^2(\Theta){\rm
d}x=\frac{\exp({-K[(1+{\displaystyle\frac{t}{1.15d\Theta^2}})\frac{\Theta}{60}]^2})}{t+1.15d\Theta^2}{\rm
d}t{\rm d}\Theta
\end{equation}

For a viable point source with light-curve $I(\tau)$, the
normalized light-curve in the halo is
\begin{equation}\label{ddsj}
B(t,\Theta)=\frac{\int^t_{t-t_m}I(\tau)H(t-\tau,\Theta)d\tau}{\int^t_{t-t_m}H(t-\tau,\Theta)d\tau}
\end{equation}
where $t_m(\Theta)$ is the possible maximal time delay.

In TS1973, $I(\tau)$ and $B(t,\Theta)$ are compared directly to determine
distance.

Suppose the light-curve $I(t)$ of the point source and the
light-curve $B(t,\omega)$ of the halo at $\omega$ radius are
stationary random signals, $i(\omega)$ and $b(\omega,\Theta)$ are
their Fourier transformation respectively. It is easy to show that
\begin{equation}
b(\omega,\Theta)=i(\omega)h(\omega,\Theta)
\end{equation}
where
\begin{equation}
h(\omega,\Theta)=\int_0^{t_m}H(t,\Theta)e^{-i\omega t}{\rm
d}t\bigg/\int_{0}^{t_m}H(t,\Theta){\rm d}t
\end{equation}
is the normalized integral transformation of the signal arrival
time distribution function $H(t,\Theta)$.

Denote the power density spectrum of the point source and the halo
as $P_b(\omega,\Theta)$ and $P_i(\omega)$, we get
\begin{equation}
P_b(\omega,\Theta)=P_i(\omega)h^2(\omega,\Theta)
\end{equation}

Given the power density spectra of the point source
and the halo at different radius, $h(\omega,\Theta)$ and
$H(\omega,\Theta)$ are easy to know, and $d$ can thus be determined by
fitting equation (3).

\section{Application to Chandra data on Cyg X-3}

\begin{figure}[htb]
\begin{center}
\vskip 1.8in \includegraphics{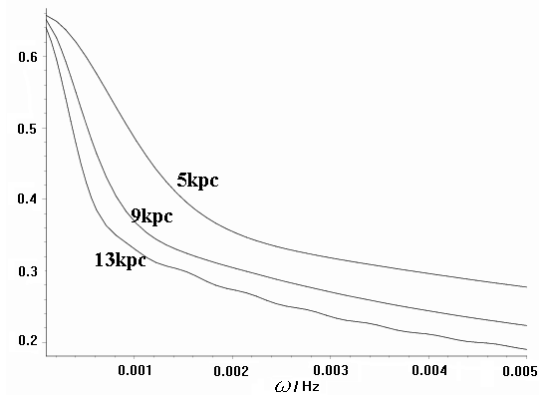} \caption {The theoretical value of the Function
$h^2(\omega,\Theta=4^{\prime\prime})/h^2(\omega,\Theta=2^{\prime\prime})$ at 5.0-7.0 energy
band. The three curves take $d=5,9,13$ kpc respectively.}
\end{center}
\end{figure}

Now we apply this method to the Chandra data of X-ray binary Cyg
X-3.  Cyg X-3 is an eclipsing X-ray binary with an orbital period
of 4.8 hours (Brinkman et al. 1972). Cyg X-3 was observed with the
High Energy Transmission Grating Spectrometer (HETGS) on board of
Chandra on October 20, 1999, with a total time of 12.3 ksec,
starting at 01:11:38 UT. The observation covered the binary phases
from -0.3 to +0.4. We only use the zeroth order data in this
study.

Because the image of the point source is severely damaged by the
pileup effect, we can only campare the power density spectra at
different and radius. The theoretical difference of the power
density spectrum of the halo at 2 arcsec and 4 arcsec in 5.0-7.0
keV energy band are shown in Fig. 2. Then we calculate
$h^2(\omega,\Theta=[4^{\prime\prime},6^{\prime\prime}])/h^2(\omega,\Theta=[2^{\prime\prime},4^{\prime\prime}])$ in 5.0-7.0
keV energy band, the result shows in Fig. 3. Because the error bar
of the data in Fig. 3 is very big, it is hard to fit the data with
theoretical curves. For simplicity, we fit the data with a
exponential curve. The turning point of the fitting curve lies at
about 0.001 Hz, and the corresponding value of $d$ is 9 kpc.
According to Fig. 3, it seems the real turning point should small
than 0.002 Hz, so we estimate $d$ is at least 5 kpc.

\begin{figure}[htb]
\begin{center}
\vskip 1.7in \includegraphics{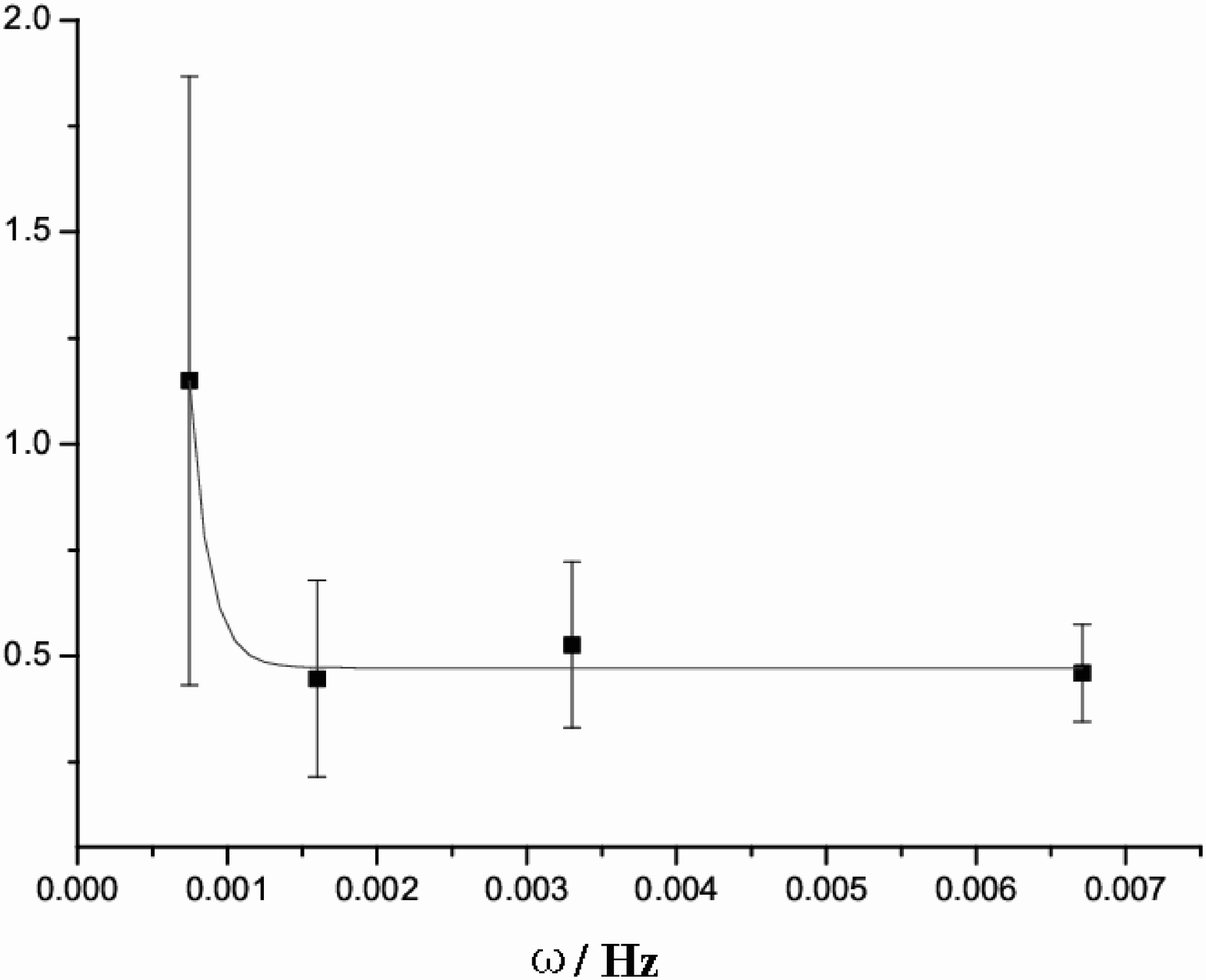} \caption {Observed value of the function
$h^2(\omega,\Theta=[4^{\prime\prime},6^{\prime\prime}])/h^2(\omega,\Theta=[2^{\prime\prime},4^{\prime\prime}])$ in 5.0-7.0
keV energy band.}
\end{center}
\end{figure}

\section{Discussion}

Our result is consistent with previous estimate on the distance of Cyg X-3. Predehl et al.
(2000) used TS1973 method and obtained $9^{+4}_{-2}$ kpc, Dickey (1983)
found a lower limit of 9.2 kpc 1 using 21 cm wavelength absorption
data, Predehl \& Schmitt (1995) derived 8 kpc as distance through
the galactic dust layer from their comparison of X-ray scattering
and absorption.

Our method and TS1973 method both determine the geometric distance
instead of physical distance. The physical distance depend on the
estimation of many physical parameters, i.e., absorption optical
depth. The uncertainties of the geometric distance determination
come mainly from the dust density distribution along the line of
sight. In principle, it can be solved by carefully fitting the
shape and strength of the halo. In TS1973 method the X-ray source
must have relatively obvious variations in the time interval
comparable to the delay time (see Fig. 4. in Predehl et al. 2000).
But in our method this requirement is not needed.

Although the Chandra Observatory has high resolution, its
effective area is relatively small, therefore the count rate of the Cyg X-3
halo is not high enough for good statistical study with our method. The future X-ray satellites such as
{\it Constellation-X} will provide us with high count rate data, and our
method can be applied to many more galactic variable X-ray sources.

{\bf Acknowledgement: }This study
is supported in part by the Special Funds for Major State Basic Research Projects and by the National
Natural Science Foundation of China. SNZ also acknowledges supports
by NASA's Marshall Space Flight Center and through NASA's Long Term Space Astrophysics Program.

\end{document}